\documentclass[twocolumn,showpacs,aps]{revtex4}
\usepackage{graphicx}
\usepackage{dcolumn}
\begin{document}
\title{
Multiple anion formation in low energy electron collisions with Tl atoms: 
Regge-pole prediction}
\author{A.\ Z.\ Msezane}
\affiliation{Center for Theoretical Studies of
        Physical Systems, Clark Atlanta University, Atlanta, Georgia 30314 USA}
\author{Z.\ Felfli}
\affiliation{Center for Theoretical Studies of
        Physical Systems, Clark Atlanta University, Atlanta, Georgia 30314 USA}
\author{D.\ Sokolovski}
\affiliation{School of Mathematics and Physics,
     Queen's University of Belfast,
                  Belfast, BT7 \#1NN, United Kingdom}
                  \date{\today}
\abstract
The complex angular momentum (CAM) calculated low-energy 0 $\leq$ E $\leq$ 5eV£  
electron elastic total cross section (TCS) for In is benchmarked 
through its recently measured electron affinity (Walter et {\it et al}, 
Phys. Rev. A {\bf 82}, 032507 (2010)). The TCSs for Tl and Ga atoms are 
then calculated using the CAM method.  From the dramatically sharp 
resonances in the TCSs binding energies for Tl$^-$and Ga$^-$ negative 
ions formed during the collisions are extracted and compared with 
existing values.  Three stable bound negative ions of Tl are predicted.}
\pacs{34.80.Bm; 32.10.Hq}
\maketitle
Recently, Walter {\it et al.} [1] have measured using infrared 
photodetachment threshold spectroscopy the electron affinity (EA) of 
In to be 383.92(6) meV.  This value, important for benchmarking theory, 
compares very well with most theoretical EAs [2-7] but differs 
substantially from previous measurements [8, 9].  For the Tl atom 
the calculated EAs [3-5] differ significantly from the measured ones 
[8, 10] while for Ga the agreement among the theoretical EAs is generally 
good, but the theoretical EAs [3-6] deviate   substantially from the 
experimental values [8, 11]. The recently observed excellent 
catalytic properties of Au and Pd nanoparticles and the exceptional 
catalytic activity of the Au-Pd catalyst when catalyzing H$_2$2O$_2$ [12] 
have provided a new impetus to study low-energy electron elastic 
scattering from atoms in general, in search of nanocatalysts [13].  
To our knowledge, there are no electron scattering cross sections 
for the In, Tl and Ga atoms available in the literature within the 
electron impact energy range of interest in the present work.

In this paper we explore low-energy  E $<$ 5.0eV elastic collisions 
between an electron and the complex atoms In, Tl and Ga through the 
calculation of the elastic total cross sections (TCSs) and search 
for long-lived resonances.  These, if they exist, are manifestations 
of the formation of stable weakly bound ground and excited negative 
ions as resonances [14, 15].  The choice of Tl and Ga is based on 
the fact that these are isoelectronic to In and may help in 
understanding the behavior of electronic affinities along isoelectronic 
sequences. From the energy positions of the characteristic resonances 
we extract the binding energies (BEs) of the ground and the excited 
negative ions formed during the collisions.  The recent 
complex angular momentum (CAM) or Regge-pole methodology [16, 17] is 
used in the investigations; it requires no {\it a priori} knowledge of the 
experimental or other theoretical data as inputs.  The imaginary part 
of the CAM, Im L, is used to distinguish between the shape resonances 
(short-lived resonances) and the stable bound states of the negative 
ions (long-lived resonances) formed as Regge resonances in the 
electron-atom collisions.

Crucial to the existence and stability of most negative ions are the 
mechanisms of electron-electron correlations and core-polarization 
interactions.  In the CAM description of scattering we use the 
Mulholland formula wherein is embedded the former effects in the 
form [16, 17] (atomic units are used throughout)
\begin{eqnarray}\label{1.1}
\sigma_{tot}(E)=4\pi k^{-2} \int_0^\infty Re[1-S(\lambda)]\lambda d\lambda
-8\pi^2 k^{-2}
\nonumber
\\
\sum_n  Im\frac{\lambda_n \rho_n }{1+\exp(-2\pi i \lambda_n)}
+I(E)
\end{eqnarray}
where $S$ is the scattering matrix, $k = \sqrt{(2mE)}$, with $m$ being the mass,
$\rho_n$ the residue of the S-matrix at the $n^{th}$ pole, $\lambda_n$ and
I({\it E}) contains the contributions from the integrals along the imaginary
$\lambda$-axis.  contribution has been demonstrated to be negligible [21].  
We will consider the case for which Im $\lambda_n \ll 1$ so that for constructive 
addition, Re $\lambda_n \approx 1/2, 3/2, 5/2 \cdots$, yielding $l=Re {\it L} 
\cong 0,1,2 \cdots$.  The importance of Eq. (1) is that a resonance is likely to
affect the elastic TCS when its Regge pole position is close to a 
real integer [17].
                                                                               
The calculation of the elastic TCSs and the Mulholland partial cross 
sections uses the Thomas-Fermi (T-F) type model potential in the well 
investigated form [22]
\begin{equation}\label{2}
U(r)=\frac{-Z}{r(1+aZ^{1/3}r)(1+bZ^{2/3}r^2)},
\end{equation}
where $Z$ is the nuclear charge and $a$ and $b$ are adjustable parameters.  
For small $r$, the potential describes the Coulomb attraction between 
an electron and a nucleus,$U(r) \sim -Z/r$ , while at large distances it 
mimics the polarization potential,  $U(r) \sim -1/(abr^4)$ and accounts 
properly for the vital core-polarization interaction at very low energies.  
The effective potential
\begin{equation}\label{3}
V(r) = U(r) + \frac{L(L+1)}{2r^2}
\end{equation}
is considered here as a continuous function of the variables  and L.  
The potential, Eq. (2) has been used successfully with the appropriate 
values of $a$ and $b$.  When the TCS as a function of $b$ has a 
resonance [21] corresponding to the formation of a stable bound 
negative ion, this resonance is longest lived for a given value of 
the energy which corresponds to the electron affinity of the system 
(for ground state collisions).  This was found to be the case for 
all the systems we have investigated thus far.  This fixes the 
optimal value of $b$ for Eq. (2).  The optimal value of $a$
was found to be 0.2 for the three atoms considered here. 
In the study of low-energy electron scattering from Cu atoms, it was 
demonstrated that the ground and excited states are polarized differently 
[23] as expected.  This explains the use in this paper of different 
values for the optimal parameter $b$ for the ground and excited atoms.

The calculation of the TCSs and the Mulholland partial cross sections is 
described in [21].  Briefly, two independent approaches are adopted.  
The first integrates numerically the radial Schr\"odinger equation for 
real integer $l=Re L$ values of L to sufficiently large $r$ values.  The 
$S$-matrix is then obtained and the TCSs are evaluated as the traditional 
sum over partial waves, with the index of summation being $l$.  The second 
part calculates the poles positions and residues of the $S$-matrix, 
$S(L,k)$, following a method similar to that of Burke and Tate [24]. 
In the method the two linearly independent solutions, $f_L$ and $g_L$, 
of the Schr\"odinger equation are evaluated as Bessel functions of 
complex order and the $S$-matrix, which is defined by the asymptotic 
boundary condition of the solution of the Schr\"odinger equation, is 
thus evaluated.  Further details of the calculation may be found in [24].

$Im L$ is important in distinguishing between the shape resonances 
(short-lived resonances) and the stable bound, both ground and excited, 
states of the negative ions (long-lived resonances) formed as Regge 
resonances in the electron-atom scattering [21]. In the definitions 
of Connor [25] and the applications [21] the physical interpretation 
of $Im L$ is given.  It corresponds inversely to the angular life of the 
complex formed during the collision.  A small $Im L$ implies that 
the system orbits many times before decaying, while a large $Im L$ 
value denotes a short-lived state.  For a true bound state, namely 
$E < 0$, $Im L=0$ and therefore the angular life, $1/[Im L] \rightarrow 
\infty$, implying that the system can never decay.  $Im L$ is also used 
to differentiate subtleties between the bound and the excited states 
of the negative ions formed as resonances during the collisions.

\section{Results}
\label{sec:2}
Figure 1 presents the elastic TCS for In.  Near threshold the curve 
is characterized by a Ramsauer-Townsend (R-T) minimum at 0.0662 eV 
and a shape resonance at 0.236 eV.  Immediately following the shape 
resonance, the very sharp resonance at 0.380 eV corresponds to the 
negative ion formed during the collision of the electron and the ground 
state In atom and defines the EA of In.  The value compares excellently 
with the latest measurement [1] and calculation [4]. The TCS curve 
typifies many such TCSs that have already been calculated, such as 
those of the lanthanide atoms.
                                                                              
Thus, the complex angular momentum calculated low-energy 0 $\leq$ E 
$<$ 0.7eV electron elastic total cross section for In is benchmarked 
to the recent measurement [1] through the electron affinity.  Henceforth 
the CAM method will be used to calculate the electron elastic scattering 
cross sections for Tl and Ga.  From the TCSs the binding energies (BEs) 
of the resultant negative ions of Tl$^-$ and Ga$^-$ formed during the 
collisions as resonances will be extracted and compared with existing values.

\begin{figure}
\includegraphics{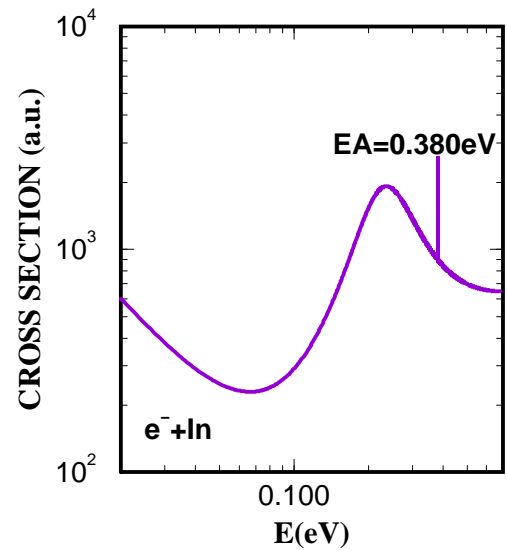}
\caption{
Elastic TCS ({\it a.u.}) for In atoms showing the R-T minimum and the shape 
resonance followed by the dramatic resonance at 0.380 eV, corresponding 
to the EA of In.}
\end{figure}
Figure 2 contrasts the low-energy 0 $\leq$ E $<$ 5eV electron-Tl elastic
scattering TCSs for the ground state, curve (a) and excited states, 
curves (b) and (c).  The structure of each curve is significantly 
different from that of the other. This is indicative of the importance 
of the electron-electron correlations and core-polarization interactions 
in the electron-Tl scattering, at both the ground and the excited states 
levels.  In the energy region of the structures, the ground state cross 
section is characterized by a R-T minimum at 0.733 eV followed by a 
shape resonance at 1.141 eV and then by a deeper and broader second 
minimum at about 2.193 eV.  The very sharp resonance right in the minimum 
corresponds to the stable bound state of the Tl$^-$ negative ion formed 
during the collision as a Regge resonance and determines the EA of Tl; 
its value is 2.415 eV.  Most significant here is that the EA of Tl 
is very close to those of Au and Pt [26] and its TCS resembles those 
of Au and Pt as well. This configuration of resonances and minima in 
the elastic TCS, typified by those of the Au and Pt TCSs, represents 
a signature of good nanocatalysts [27].  Perhaps, Tl can replace Au or 
Pt as a possible nanocatalyst in some situations and reduce the costs 
significantly. This calls for immediate experimental investigation.

Curves (b) and (c) represent electron scattering TCSs for excited Tl 
atoms, resulting in the formation of Tl$^-$ negative ions.  The sharp 
curves with binding energies (BEs) of 0.281 eV and 0.0664 eV correspond 
respectively to Tl$^-$ ions in their first and second excited states.   
A very important revelation in the comparison is the appearance of the 
bound state resonances of the negative ions together with the shape 
resonances of the ground and the excited states. Both theoretical 
calculations and experimental measurements could easily mistake one 
for the other.  This could also be problematic in the use of the Wigner 
threshold law in high precision measurements of BEs of valence electrons 
using photodetachment threshold spectroscopy.  Furthermore, the determination 
of the R-T minimum of the ground state could be hindered since it is 
mixed in together with the cross sections for the excited states.

%
Indeed, the misidentification is evident in the comparison of the 
available theoretical and experimental EAs of Tl, presented in Table II.  
For Tl the EA values of 0.27 eV [3] and 0.291 eV [4] compare excellently 
with our calculated binding energy of 0.281 eV.  Since our EA for Tl 
is 2.415 eV, we conclude that these theoretical values correspond to 
the BE of an excited Tl$^-$ anion and not to the EA as claimed. So, 
the EA values of Tl reported by the various calculations and measurements 
presented in Table II do not correspond to the EA; they are the BEs 
of the first excited state of the Tl$^-$ anion.  The various calculations 
agree reasonably well with one another and with the experiment [10], 
although it has a large error margin.
\begin{figure}
\includegraphics{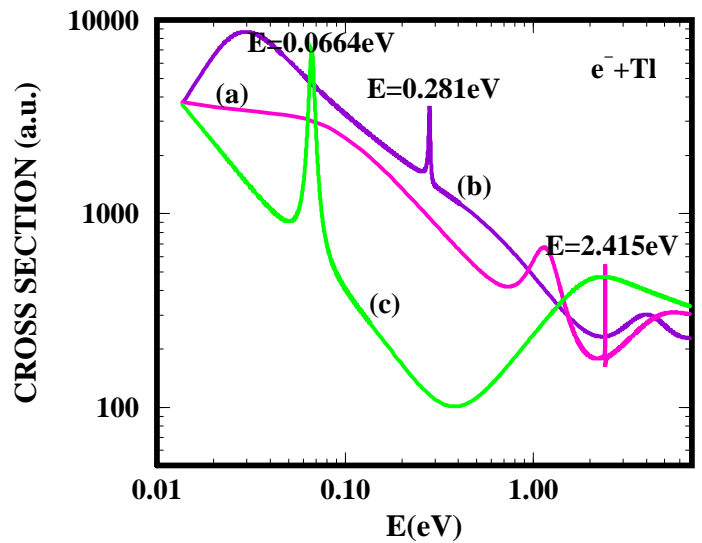}
\caption{
Total cross sections ({\it a.u.}) for electron elastic scattering from Tl atoms 
versus E (eV), are contrasted.   The curves (a), (b) and (c) represent 
respectively the ground state, first excited state and second excited state.  
All the curves are characterized by very sharp resonance structures corresponding 
to the formation of Tl$^-$ negative ions during the collisions.  Note that 
for the ground state curve the position of the bound state of the Tl$^-$ 
anion is at the second minimum.}
\end{figure}

In figure 3 the TCS for the electron-Ga scattering is presented. This 
curve resembles that of the first excited state TCS for Tl.  It is 
characterized by the usual shape resonance, followed by a dramatically 
sharp resonance which corresponds to the bound state of the Ga$^-$ 
negative ion formed during the collision.  The BE of the negative ion 
is determined to be 0.222 eV which can be compared with the data in 
Table II.  Ref. [3] has the EA of Ga as 0.29 eV, while the experiment [8] 
has the value 0.30$\pm$0.15eV.  Both our value and the other theoretical 
one [3] agree reasonably well with each other and with the experiment, 
although it has a large error margin.
\begin{figure}
\includegraphics{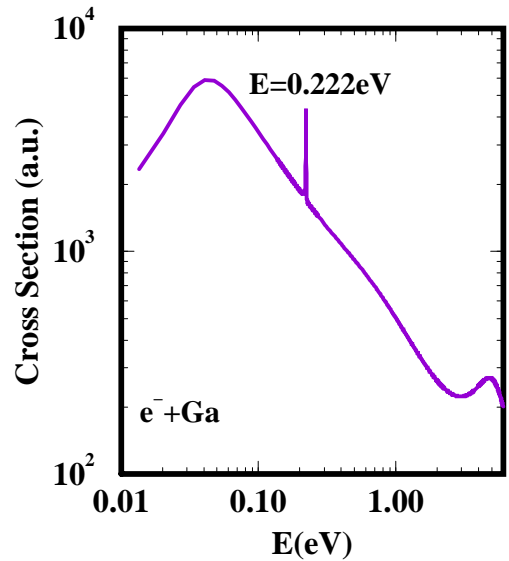}
\caption{
Elastic TCS ({\it a.u.}) for Ga atoms showing the shape resonance followed by the 
dramatic resonance, corresponding to a bound state of the Ga$^-$ anion.
}
\end{figure}

Our extracted BEs from the resonances in the TCSs of In, Tl and Ga 
atoms are tabulated in Table I and in Table II where they are compared 
with other theoretical calculations and measurements.
                                                                              
\begin{table}[tbh]
\caption{
Calculated binding energies, BEs (eV), shape resonances, SRs (eV) 
and minima, $1^{st}$ and $2^{nd}$ min. (eV) for In, Tl and Ga atoms.
}
\label{table I}
\begin{tabular}{|c|c|c|c|c|c|c|} 
\hline \hline
Z & Atom & State &  1st min. & SR & 2nd min. &  BE  \\
\hline
49 &  In & ground &  0.0662 & 0.236 & ---  & 0.380 \\
\hline
81 & Tl & ground &  0.733 & 1.141 & 2.193 & 2.415  \\
& & 1$^{st}$ excited &  --- & 0.0295 & --- & 0.281   \\
& & 2$^{nd}$ excited &  0.503 & --- & --- & 0.0664   \\
\hline
31 & Ga & ground &  --- & 0.0407 & --- & 0.222 \\
\hline \hline
\end{tabular}
\end{table}
\begin{small}
\begin{table}[tbh]
\caption{
Measured and calculated EAs (eV) for In, Tl and Ga are compared with 
the present calculated binding energies, BEs (eV).}
\label{table II}
\begin{tabular}{|c|c|c|c|c|} 
\hline \hline
Z & Atom & EA, expt. & EA, theory& BE, this work \\
49 & In & 0.38392(6) [1] & 0.371 [2] & 0.380 \\
& & 0.30 $\pm$ 0.20 [8] & 0.380 [3] &  \\
& & 0.404(9) [9] & 0.393 [4] & \\
& & & 0.419 [5] & \\
& & & 0.374 [6] & \\
& & & 0.403 [7] & \\
\hline
81 & Tl & 0.377(13) [10] & 0.27 [3] & 0.0664, 0.281, 2.415 \\
& & 0.20 $\pm$ 0.20 [8] &  0.291 [4] & \\
& & & 0.40 $\pm$ 0.05 [5] & \\
\hline
31 & Ga & 0.30 $\pm$ 0.15[8] & 0.29  [3] &  0.222 \\
& & 0.43(3) [11]& 0.305 [4] & \\
 & & & 0.301 [5]a & \\
 & & & 0.297(13) [6] & \\
\hline \hline
\end{tabular}
\end{table}
\end{small}

We have benchmarked the CAM calculated TCS for the electron-In scattering 
through the recently measured EA [1].  Our EA agrees excellently with that 
of the measurement and the calculated value [3].  The CAM method has then 
been used to evaluate the TCSs for the electron scattering from Tl and Ga 
atoms.  Binding energies for electron attachment to the ground state of Tl 
and when the Tl atom is excited have been extracted from the TCSs and 
compared with the available data.  Our calculated BE for the first excited 
state of the Tl$^-$ negative ion agrees excellently with the EAs of Refs. 
[3, 4]  and reasonably well with the EA measured by [10].  However, our 
calculated EA for the Tl atom is 2.415 eV.  Consequently, the theoretical 
and the experimental EAs for Tl in the published literature actually 
correspond to the BE of the first excited state of the Tl$^-$ negative ion.  
This calls for immediate experimental verification.

We also conclude from the configuration of the resonances and minima 
in the TCSs for Tl that Tl promises to be a good nanocatalyst (see Ref. 
[27] for discussions), capable of replacing Au and/or Pt in some 
applications.  This also calls for experimental investigation.  Finally, 
we predict the formation as resonances of three stable bound states of 
the Tl$^-$ negative ion during the collision of a slow electron with Tl 
atoms; the TCSs for Tl are similar to those of the Ag atom [27].

Work supported by U. S. DOE, Division of Chemical Sciences, Office of Basic 
Energy Sciences, Office of Energy Research and the NSF funded CAU CFNM.
The computing facilities at the Queen's University of Belfast, UK and of DOE 
Office of Science, NERSC are reatly appreciated.

{}
\end{document}